\documentstyle [aps,multicol,prl]{revtex}
\begin{document}                                                    
\draft                                                      
\title{ Mean-Field Approach to Charge, Orbital, and Spin Ordering in 
Manganites}
\author{Sudhakar Yarlagadda and C. S. Ting }
\address{
Texas Center for Superconductivity, 
University of Houston, TX 77204-5932}
\date{\today}
\maketitle

\begin{abstract}
We present a mean-field theory of charge, orbital, and spin ordering in 
manganites at 50\% and 0\% dopings by considering
Jahn-Teller interaction, nearest-neighbor repulsion, and 
no single-site double occupancy.
 For spinless fermions, we show that
Jahn-Teller
 distortion and charge-orbital ordering
 occur simultaneously.
In  our two-dimensional model at 50\% doping,
for small nearest-neighbor repulsion
 the system is orbitally polarized while for larger repulsion
the system undergoes $CE$ type ordering.
As for the 0\% doping case,
the ground state is orbitally antiferromagnetic.
Upon including spin degree of freedom,
 at both 50\% and 0\% dopings 
the ordering remains the same at small 
 antiferromagnetic coupling between adjacent core spins.
\end{abstract}
\pacs{71.45.Lr, 71.30.+h, 71.38.+i, 75.10.-b  }

\nopagebreak
\begin{multicols}{2}
                                                                  
Manganites of the form $A_{1-x} B_{x} Mn O_3$ (with A=La, Pr, etc; 
B=Ca, Sr, etc) display a rich variety
of phenomena and phases as the doping $x$ is varied 
\cite {ramirez}.
In particular close to 0\%  doping
the systems display 
  orbital ordering at higher temperatures while at lower
temperatures magnetic ordering results \cite{rodriguez}. 
In narrow band materials like
$Pr_{0.5} Ca_{0.5} Mn O_3$, as the temperature is lowered
to $T_{CO}$ the system undergoes $CE$ type charge-orbital (CO) ordering
and at a much lower temperature $T_N$ peculiar antiferromagnetic
(AF) order sets in \cite{tokura,zimmermann1} (see Fig. \ref{fig1}).
In wider band materials like $Nd _{0.5} Sr_{0.5} MnO_3$
 ferromagnetic ordering occurs at higher
temperatures and at lower temperatures simultaneous charge, orbital, and
spin (AF) ordering results \cite{tokura,kawano}.
In layered-perovskites like $La_{0.5} Sr_{1.5} MnO_4$ it has
been reported that $T_{CO}$ is higher than $T_N$ \cite{murakami}.
In systems where $T_{CO}$ and $T_N$ coincide,
Solovyev and Terakura argue that the CO-ordering
is a result of the peculiar AF ordering along
with Hund's rule physics \cite{solovyev}.
 van den Brink {\em et al.} too analyze 
 systems like $Nd_{0.5} Sr_{0.5} MnO_3$ 
and find that because of a topological factor
in the hopping, the one-dimensional ferromagnetic zigzag chains 
are orbitally ordered \cite{vandenbrink}.
 However Refs. \cite{solovyev,vandenbrink} do not include
electron-phonon interaction physics nor do they
 consider narrow band systems like $Pr_{0.5} Ca_{0.5} MnO_3$.
Evidence of significant change in lattice parameters has been
reported at $T_{CO}$ for $Pr_{1-x} Ca_{x} MnO_3$ (at x=0.4, 0.5)
\cite {zimmermann2} and in $La_{0.5}Ca_{0.5} MnO_3$
 \cite{radaelli}.

  In this paper
we derive a general framework for analyzing the concomitant
 charge density wave (CDW)
and orbital density wave (ODW) instability
 by considering Jahn-Teller interactions and showing
that the two-dimensional (2D) ordering can be understood in terms
of the  one-dimensional (1D) Kohn anomaly
 at wavevector $q = 2k_F$.
The problem essentially reduces to comparing the energy
resulting from the hopping term for various possible ordered
states. We show that 
Jahn-Teller
 distortion is accompanied by 
CO-ordering for the carriers.
We find that above a critical value of nearest-neighbor repulsion
and for small antiferromagnetic coupling between adjacent core spins,
$CE$ type charge, orbital, and spin (COS)
ordering is realized at 50\% doping.
At 0 \% doping orbitally antiferromagnetic order results.

 We first study the ordering phenomena without spin degree
of freedom because $T_{CO} \ge T_N$.
We begin by considering CDW instability for a 1D
Holstein model given below:
 \begin{eqnarray} 
H=
- t
\sum _{\langle i j \rangle} 
  a _{i }^{\dagger}  
  a _{j } 
  && 
+\sum_{\vec{q}} 
\omega_{0}
f_{\vec{q}}^{\dagger}
f_{\vec{q}} 
 \nonumber \\                           
  && 
+
\frac{A}{\sqrt{N}} 
 \sum_{ \vec{q}, \vec{k} }
a^{\dagger}_{\vec{k} +\vec{q}}
a_{\vec{k}}
( f_{\vec{q}} +
f_{-\vec{q}}^{\dagger}) ,
\label{1dholst}
 \end{eqnarray} 
where
$  a _{j } ~ (f_{\vec{q}})$ is the
 electron (phonon) destruction operator,
$ \langle i j \rangle$ corresponds to nearest neighbors,
$\omega_{0}$ is the optical phonon frequency ($\hbar = 1$),
$A (< 0 )$
 is the electron-phonon coupling, 
$t$ is the hopping integral, and $N$ is the number of sites.
We know that in 1D
 the non-interacting polarizability
 $\chi_0 (\vec{q})$
(Lindhard polarizability) diverges at wavevector $q=2k_F$.
Then even in the weak-coupling limit we will have a CDW instability
because the expectation value 
 $\langle \sum_{ \vec{k} }
a^{\dagger}_{\vec{k} +\vec{q}}
a_{\vec{k}}\rangle $ diverges in the interaction part of the
Hamiltonian. Furthermore, upon calculating the double derivative
with respect to time of the ionic position coordinate $Q_{\vec{q}}
=\sqrt{1/(2M\omega_0)} (f_{\vec{q}} + f_{-\vec{q}}^{\dagger})$ by
using the relation $\ddot{Q}_{\vec{q}} = - [[Q_{\vec{q}}, H], H ]$
we get  
$\ddot{Q}_{\vec{q}} =
- \omega _0^2 Q_{\vec{q}}
 - 2 A^2 \omega_0 \chi_0 (\vec{q}) Q_{\vec{q}}$.
Thus we see that the renormalized phonon frequency becomes soft even
for vanishingly small electron-phonon interaction leading to lattice
distortion. The order parameter for the
 CDW state is $\langle f_j \rangle
= \Delta _j e^{i \phi _j}$.
Then within mean-field we get from Eq. (\ref{1dholst}),
 after averaging over phonon coordinates,
 the following (see Ref. \cite{gruner}
for CDW in metals):
 \begin{eqnarray} 
\bar{H}= - t
\sum _{\langle i j \rangle} 
  a _{i }^{\dagger}  a _{j } 
+\sum_{j} \omega_{0} \Delta _{j} ^2
+
2 A \sum_{ j }
a^{\dagger}_{j}
a_{j} \Delta _{j} \cos \phi_j .
\label{1dldw}
 \end{eqnarray} 
The energy in the above equation is minimized for $\phi _j =0 $
and for
$
\omega_0 \Delta_j =
-A\langle a^{\dagger}_{j} a_j \rangle$.

We will now solve for the CDW state at half-filling,
using $\Delta_j = \Delta_0 /2 + \Delta_1 \cos (\pi j )$. We first
note that $\Delta_0 =-A/\omega_0 $ and that $\Delta_1 \leq \Delta_0/2$.
Next we add the nearest-neighbor interaction term 
$V_{NN} =V \sum_{j, \delta} n_j n_{j+\delta} $ with $\delta$
 being nearest-neighbors
and  make the mean-field  approximation
$n_j n_{j+1} = n_j \langle n_{j+1} \rangle + n_{j+1} \langle n_j \rangle
- \langle n_{j+1} \rangle \langle n_j \rangle$.
Then the Hamiltonian in momentum space,
 on folding the Brillouin zone, gets modified to be
 \begin{eqnarray} 
  \bar{H}=
 &&
\sum _{k} [H_k + H_{k+ \pi} 
+2A B_+\Delta_1 
( a _{k }^{\dagger}  a _{k + \pi } + H.c.) ]
\nonumber \\
  && 
  +
\omega_0 N (B_- \Delta_0^2 /4 + B_+ \Delta_1 ^2 ) ,
\label{1dcdwV}
 \end{eqnarray} 
where $H_k \equiv (-2t \cos k + A B_- \Delta_0 )
 a _{k }^{\dagger}  a _{k }$, 
 $B_{\pm} = 1 \pm zV \omega_{0}/A^{2}$ 
with $z$  being the coordination number,
and 
the reduced Brillouin zone is $-\pi/2 \le k \le \pi/2$.
Then diagonalizing the above Hamiltonian yields the
 eigen energies
$E_{k}^{\pm} = AB_{-} \Delta_0
 \pm \sqrt{ 4 t^2 \cos^2 k + 4 A^2 B^2_{+} \Delta_1^2}$.
Upon minimizing the total energy with respect to $\Delta_1$
we get
 the optimum  $\Delta_1 \approx - A/(2 \omega_0 ) 
\left [ 1 - (\lambda / B_+ )^2 + 1/4
 (\lambda/ B_+)^4 \right ]$ 
  for small values of 
$\lambda/ B_+$ where the polaron size parameter
 $\lambda \equiv t \omega_0 /A^2 $.
Then the minimized energy per particle is $-A^2 /\omega_0 \left [ 1 +
\lambda ^2 /B_{+} - \lambda ^4/(4 B_{+}^3) \right ] $
 where the last two terms involving $\lambda$ represent
 the {\em hopping term energy}.
 We see that including nearest-neighbor interactions
does not alter the results qualitatively although it {\em enhances} the
charge modulation.

We will now proceed to consider manganite systems with two orbitals per
site. The Hamiltonian consists of three
parts $H = H_1 + H_2 + H_3$ where $H_1$ is the hopping term,
$H_2$ the ionic term, and $H_3$ the electron-ion interaction term.
The hopping term is given by $-t \sum _{\langle ij \rangle}
e^{\dagger}_i e_j$ where 
 $\langle ij \rangle$ corresponds
to  nearest-neighbors in the
$\alpha (=x,y,$ or $ z$) direction and $e$
 is the destruction operator of the
orbital $\psi_{3\alpha ^2 - r^2}$. 
Here it should be mentioned that for propagation in the x-direction
the orbital basis is
 $\psi_{3x^2 -r^2}$ and $\psi_{y^2-z^2}$, for
y-direction it is $\psi_{3y^2-r^2}$ and $\psi_{z^2-x^2}$, etc.

To analyze the ordering problem,
we consider
the orthonormal wavefunctions
 $\psi_{X[Y]}
= \frac{1}{\sqrt{6}}[(\sqrt{3} + 1) \psi_{3x^2[y^2] -r^2} 
+ (\sqrt{3} -1 ) 
\psi_{3y^2[x^2]- r^2} ]$ the justification for which will be given
below \cite{lisheng}.
 The interaction term in this basis is  given by 
\begin{eqnarray}
H_3 = 
A \sqrt{2 M \omega_0} \sum_{j} [ &&
Q_{2j}(a^{\dagger}_{Xj} a_{Xj} -  a^{\dagger}_{Yj} a_{ Y j} )
\nonumber \\
&&
 +
Q_{3j}(a^{\dagger}_{Xj} a_{Yj} +  a^{\dagger}_{Yj} a_{ X j} ) ] ,
\end{eqnarray}
where $a_{X j}$ and $a_{Y j}$ are the 
destruction operators of the electrons in
$\psi_{X}$ and $\psi_{Y}$ orbitals at site $j$,
with $Q_2$ and $Q_3$ corresponding to the
two normal modes. 
The ionic part of the Hamiltonian is then given by
\begin{equation}
H_{2} = \frac{K}{2} \sum_{j} (Q_{2j}^2 + Q_{3j}^2) + \frac{M}{2}
\sum _{j} (\dot{Q}_{2j}^2 +\dot{Q}_{3j}^2 ) .
\end{equation}
We further enforce that at most only one orbital
can be occupied at each site by setting 
$\langle a^{\dagger}_{X j} a_{X j} \rangle
\langle a^{\dagger}_{Y j} a_{Y j} \rangle =0$.

We study the problem in 2D
(say the xy-plane) and
observe
 that the overlap between 
 $\psi_{X}$ and $\psi_{3x^2 - r^2}$ is greater than $96.6 \%$
and
 $ \langle \psi_X \psi_{3x^2-r^2}\rangle =
 \langle \psi_Y \psi_{y^2-z^2}\rangle 
=\langle \psi_Y \psi_{3y^2-r^2}\rangle 
=| \langle \psi_X \psi_{z^2-x^2}\rangle|$.
Hence based on the hopping term of
the Hamiltonian, to bring out the essential 
 physics,
 we approximate $H_1$ to be
\begin{equation}
H_1 \approx -2t \sum_{\vec{k}} \left ( \cos k_x a^{\dagger}_{X\vec{k}}
 a_{X\vec{k}} +
\cos k_y 
 a^{\dagger}_{Y\vec{k}} a_{Y \vec{k}} \right )  .
\label{approxke}
\end{equation}
The 
$50 \%$ doping case will be first considered.
In $H_3$ we 
 note that 
 $\langle  \sum_{\vec{k}}
a^{\dagger}_{X\vec{k} + \vec{q}}a_{X\vec{k}} -
a^{\dagger}_{Y\vec{k} + \vec{q}}a_{Y\vec{k}} \rangle$
 diverges
at $\vec{q}= \vec{Q}/2 = [\pi/2 , \pi/2]$ for the orbitally
 unpolarized case
and at $\vec{q} = [\pi,...]$ or $\vec{q} = [...,\pi]$
 for the orbitally polarized case
leading to a concomitant CDW and ODW instability with the phonon
frequency corresponding to $Q_{2 \vec{q}}$ going soft.
To see this transparently
we first recognize that for the electrons in the $\psi_{X}$ ($\psi_Y$)
orbitals
the dispersion relation is like the 1D case due to which
 the Fermi sea is rectangular with $-\pi/4 \leq k_x (k_y) \leq \pi/4$
and $-\pi \leq k_y (k_x) \leq \pi$ for the orbitally unpolarized
case.
 Hence, the Lindhard polarizability
diverges at $[2k_{xF},...]$ and $[..., 2k_{yF}]$ for
$\psi_{X}$ and $\psi_Y$ electrons respectively.
As for the orbitally polarized case, on choosing
 $\psi_Y$ orbitals without
loss of generality,
the Fermi sea is given by
  $-\pi/2 \leq k_y  \leq \pi/2;$
 $-\pi \leq k_x \leq \pi$ and thus the non-interacting
polarizability is singular at $[2k_{yF},...]$.
 Furthermore 
 $\langle  \sum_{\vec{k}}
a^{\dagger}_{X\vec{k} + \vec{q}}a_{X\vec{k}} -
a^{\dagger}_{Y\vec{k} + \vec{q}}a_{Y\vec{k}} \rangle$
($\langle  \sum_{\vec{k}}
a^{\dagger}_{X\vec{k} + \vec{q}}a_{Y\vec{k}} +
a^{\dagger}_{Y\vec{k} + \vec{q}}a_{X\vec{k}} \rangle$)
is similar to the longitudinal (transverse)
spin susceptibility in a Hubbard model
 with $\psi_{X}$ and $\psi_{Y}$ orbitals standing for up and down spins.
 Thus the system, to lower its energy,
prefers $H_3$ to be diagonal in the 
$\psi_{X}$ and $\psi_{Y}$ orbital basis
with {\em only} $Q_2$ mode getting excited at all sites.
The Hamiltonian after averaging over the phonon coordinates
yields the following two-orbital 2D version of
Eq. (\ref{1dldw}) with optimum values of $\phi _j$ and $\Delta _j$:
 \begin{eqnarray} 
\bar{H}= &&
 -2t \sum_{\vec{k}} \left ( \cos k_x a^{\dagger}_{X\vec{k}}
 a_{X\vec{k}} +
\cos k_y 
 a^{\dagger}_{Y\vec{k}} a_{Y \vec{k}} \right )
 \nonumber \\                           
  && 
  -
 2 \frac{A^2}{\omega_0} \sum_{ j } 
 ( a^{\dagger}_{Xj} a_{Xj} - a^{\dagger}_{Yj} a_{Yj} ) 
\langle  
 a^{\dagger}_{Xj} a_{Xj} - a^{\dagger}_{Yj} a_{Yj} 
\rangle
 \nonumber \\                           
  && 
  +
  \frac{A^2}{\omega_0} \sum_{ j }
\langle  a^{\dagger}_{Xj} a_{Xj} - a^{\dagger}_{Yj} a_{Yj} \rangle ^2 .
\label{2dcdw}
 \end{eqnarray} 

We will now compare the energies for the 
 CDW states 
in the orbitally polarized and unpolarized states.
 For the orbitally unpolarized case
the order parameter
$\langle  a^{\dagger}_{Xj} a_{Xj} - a^{\dagger}_{Yj} a_{Yj} \rangle
= C \cos(\vec{Q}/2 \cdot \vec{R}_j)$. Then using the fact that
$\langle  a^{\dagger}_{Xj} a_{Xj}\rangle 
\langle a^{\dagger}_{Yj} a_{Yj} \rangle =0$,
 we see in the present $50 \%$ doping case
 that $|C|=1$ and that we get $CE$ type CO-ordering.
Then the CDW order parameter is given by
$\langle  a^{\dagger}_{Xj} a_{Xj} + a^{\dagger}_{Yj} a_{Yj} \rangle
=  \cos ^{2} (\vec{Q}/2 \cdot \vec{R}_j)$.
We next
introduce nearest-neighbor  interaction $V_{NN}$
and note that 
$\langle n_{j} \rangle \langle n_{j+\delta} \rangle =0 $
 and thus the CDW
is also compatible with nearest-neighbor repulsion.
Then taking $C=-1$
without loss of generality, the total Hamiltonian is
 \begin{eqnarray} 
\bar{H}=
  \sum_{\vec{k} , \alpha = X, Y}
  \sum_{n=0}^{3}
 H^{\alpha}_{\vec{k}+ n \vec{Q}/2}
+\frac{A^2 N}{2 \omega_0} ,
\label{2dham}
 \end{eqnarray} 
where
the momentum summation,
for $\psi_{X(Y)}$ electrons, is over the
 reduced Brillouin zone $-\pi /4 \le k_{x(y)}  \le \pi /4;$
 $ -\pi \le k_{y(x)} \le \pi$ and
 \begin{eqnarray} 
 H^{X(Y)}_{\vec{k}} \equiv  &&
 [-2t \cos k_x + Vz] a^{\dagger}_{X(Y)\vec{k}} a_{X(Y)\vec{k}} 
\nonumber \\
&&
-(+)A \Delta _0 [  a^{\dagger}_{X(Y)\vec{k}} 
a_{X(Y)\vec{k}+ \vec{Q}/2} + H.c. ] 
\nonumber \\
&&
-Vz/2 [  a^{\dagger}_{X(Y)\vec{k} 
+ \vec{Q}} a_{X(Y)\vec{k}} + H.c. ] .
 \end{eqnarray} 
We diagonalize the Hamiltonian in Eq. (\ref{2dham}) to obtain
 \begin{eqnarray} 
 && \eta^{4}_{X(Y)} +
4 Vz\eta^{3}_{X(Y)} +
4\eta^{2}_{X(Y)}[(V^2z^2 - A^2 \Delta_0^2)-t^2] 
\nonumber \\
&&
- 8 \eta_{X(Y)} Vz t^2 
 + 16 t^2 \sin^2(k_{x(y)}) \cos^2(k_{x(y)}) =0 ,
 \end{eqnarray} 
where
$\eta_{X(Y)} \equiv E_{\vec{k}}^{X(Y)} -2Vz$
with $E_{\vec{k}} ^{X(Y)}$ being the eigen energies
 for $\psi_{X(Y)}$ electrons. 
Treating $\lambda$
 as a small parameter
we obtain after some algebra the following expressions
for the lowest energies
 \begin{eqnarray*} 
 E_{\vec{k}}^{X(Y)} = -\frac{ A^2}{\omega_0} \left [ 2 +
\frac{\lambda ^2}{ B_{+}} - \frac{\lambda ^4 }{ 4 B_{+}^2} 
\left ( \sin^2(2k_{x(y)}) + \frac{B_{-}}{B_{+}} \right ) \right ] ,
  \end{eqnarray*}                               
where terms of order
 $t\lambda ^5$
 or higher
 are neglected. On filling up the lowest band we obtain the energy per
particle to be
 $-A^2/\omega_0 [ 1 + \lambda ^2 /B_{+} 
- \lambda ^4 /(4 B_+ ^2)(1/2 + B_{-}/B_{+})]$.

As for the orbitally polarized case, 
the order parameter
is given by $\Delta _{j} =
\Delta_0 /2  + \Delta_1 \cos (\vec{Q} \cdot \vec{R}_j )$
and the problem is similar to the 1D
 one-orbital
case \cite{sudhakar1}.
On choosing the $\psi_Y$ orbitals to be occupied,
 the reduced Brillouin zone is
$-\pi/2 \leq k_y \leq \pi/2;$ 
$-\pi \leq k_x \leq \pi $.
Then the minimized energy per particle is, as before,
 $-A^2/\omega_0 [ 1 + \lambda ^2 /B_{+} - \lambda ^4 /(4 B_+ ^3)]$.
It is of interest to note that  for small nearest-neighbor
repulsion $V$,
the energy of
 the orbitally polarized state
is lower than that of the
  orbitally unpolarized state  while for
 $Vz > A^2 /\omega_0$ it is higher than that
of the $CE$ type state.

We will now consider the $0\%$ doping case.
Here we get  an ODW instability at
$\vec{q}=\vec{Q}$ leading to an 
 orbitally antiferromagnetic state.
The reduced Brillouin zone for $\psi_{X(Y)}$ orbital is 
$-\pi/2 \le k_x (k_y) \le \pi/2 ;$ $-\pi \le k_y (k_x) \le \pi$.
The ground state energy  per particle
when $\lambda$
 is  a small parameter
is obtained to be 
$Vz -A^2 /\omega _0 [ 1 + \lambda^2 /2 
- 3 \lambda ^4/32 ] $.
As for the orbitally polarized case,
the energy per particle is $- A^2/\omega_0 +   Vz$.
Thus we see that the energy for the orbitally unpolarized
case is {\em lower} than that for the orbitally polarized case
at $0 \%$ doping.

We will now include spin degree of freedom
and analyze the magnetic ordering that results from
the CO-ordering derived above.
We consider antiferromagnetic
spin coupling between adjacent localized
core spins of the form 
$ \frac{J}{2} \sum _{j, \delta} \vec{S}_{j} \cdot \vec{S}_{j + \delta} $
($J>0$ and $S=3/2$)
 and an infinitely strong
  Hund's coupling between itinerant electrons
and localized spins.
Then in the Hamiltonian,
the kinetic energy becomes
$-t \sum _{[ i j ] } \cos (\theta_{ij} /2 ) 
(a^{\dagger} _{Xi } a_{Xj}+ a^{\dagger} _{Yi } a_{Yj})$
where $[ i j ]$ are nearest-neighbors in the x(y)-direction
for $\psi _{X(Y)}$ orbital electrons,
$\theta _{ij}$ is the
angle between adjacent
 core spins  
 $\vec{S}_{i}$
  that are treated classically.
As for the 2D
 two orbital case at $50 \%$ doping,
 for a $CE$ type magnetic
ordering we get the same
expectation value
for the kinetic energy 
 because 
$\theta_{ij} =0 $ when an electron in the $\psi_{X} (\psi_{Y})$
orbital tries to jump 
to the next unoccupied orbital in the x- (y-) direction
 (see Fig. \ref{fig1}).
The nearest-neighbor core spin coupling gives
zero value.
 The contribution from the remaining {\em spin independent} terms
in the Hamiltonian
is the same as before
 and thus the total energy per particle remains the
same for this spin order.
On the other hand if the system is totally spin antiferromagnetic,
 the kinetic energy is zero
and the core spin interaction energy per particle is $-4J S^2$.
Thus we see that if the magnitude of the
hopping term energy per particle is more than 
$4JS^2$ we get
 $CE$ type COS
 ordering in 2D \cite{sudhakar3}.
As for the case when the system is orbitally
 polarized (say of $\psi_{Y}$
orbitals), if the magnetic ordering comprises of 
 spin polarized chains in the 
y-direction that are antiferromagnetically coupled in the x-direction
(see Fig. \ref{fig2}),
the energy per particle is still the same as when the spin effects are
ignored \cite{sudhakar2}. 
On the other hand if the orbitally polarized case is completely
spin antiferromagnetic again the kinetic energy is zero
and the core spin interaction energy per particle is $-4J S^2$.
Thus here too we see that if the magnitude of the
 hopping term energy per particle is less
than $4JS^2$, AF order results (see Fig. \ref{fig3}).

As for the $0 \%$ doping case,
 when the magnitude of the
hopping term energy per particle is more than $4J S^2$,
 similar analysis as above yields
an orbitally antiferromagnetic and spin ferromagnetic state.
Otherwise, the system is spin antiferromagnetic.

The above analysis can be extended to other dopings
as well.
All in all, we feel that our analysis explains the
COS ordering in layered
compounds like $La_{0.5}Sr_{1.5}MnO_{4}$ and
narrow-band systems like $Pr_{0.5}Ca_{0.5}MnO_{3}$
where $T_{CO} > T_N$. Since we have shown that
 magnetic ordering results from the CO ordering, 
 our analysis should
also hold for wider-band systems like $Nd_{0.5}Sr_{0.5}MnO_3$.


{The authors would like to thank W. P. Su,
 G. Baskaran,
 R. Pandit, L. Sheng, J. Zhu, and Y. Murakami
 for discussions.
This work was supported
by Texas Center for Superconductivity, a grant from
Texas ARP (ARP-003652-0241-1999), and by the Robert Welch Foundation.}

\end{multicols}
\begin{figure}
\caption{$CE$ ordering in x-y plane. Exaggerated
 $Q_2$ distortion is shown
by the dash-dot line and the zig-zag ferromagnetic
 chains by dashed lines.
The $Mn^{3+}$ sites are depicted by $\Psi_{X,Y}$ orbitals and core
spins, the
$Mn^{4+}$ sites by only core spins, and the oxygen sites by 
 ${\bf \bigcirc}$.}
\label{fig1}
\end{figure}

\begin{figure}
\caption{Orbitally polarized state in x-y plane.
 Exaggerated $Q_2$ distortion is shown
by the dash-dot line and the ferromagnetic chains by dashed lines.
The $Mn^{3+}$ sites are depicted by $\Psi_{Y}$ orbitals and core
spins, the
$Mn^{4+}$ sites by only core spins, and the oxygen sites by 
 ${\bf \bigcirc}$.}
\label{fig2}
\end{figure}

\begin{figure}
\caption{Phase diagram for the 50\% doping case in 2D.}
\label{fig3}
\end{figure}


\begin{references}
\bibitem{sudhakar1}   
 For $V_{NN}=0$, the ordering can also be at $ \vec{q} = [\pi , 0]$.
\bibitem{sudhakar3}
One can expect the
observed stacking of like orbitals in the z-direction
(see Ref. \cite{tokura}) when antiferromagnetic
coupling energy compensates for the nearest-neighbor repulsion in this
direction.
\end{references}
  
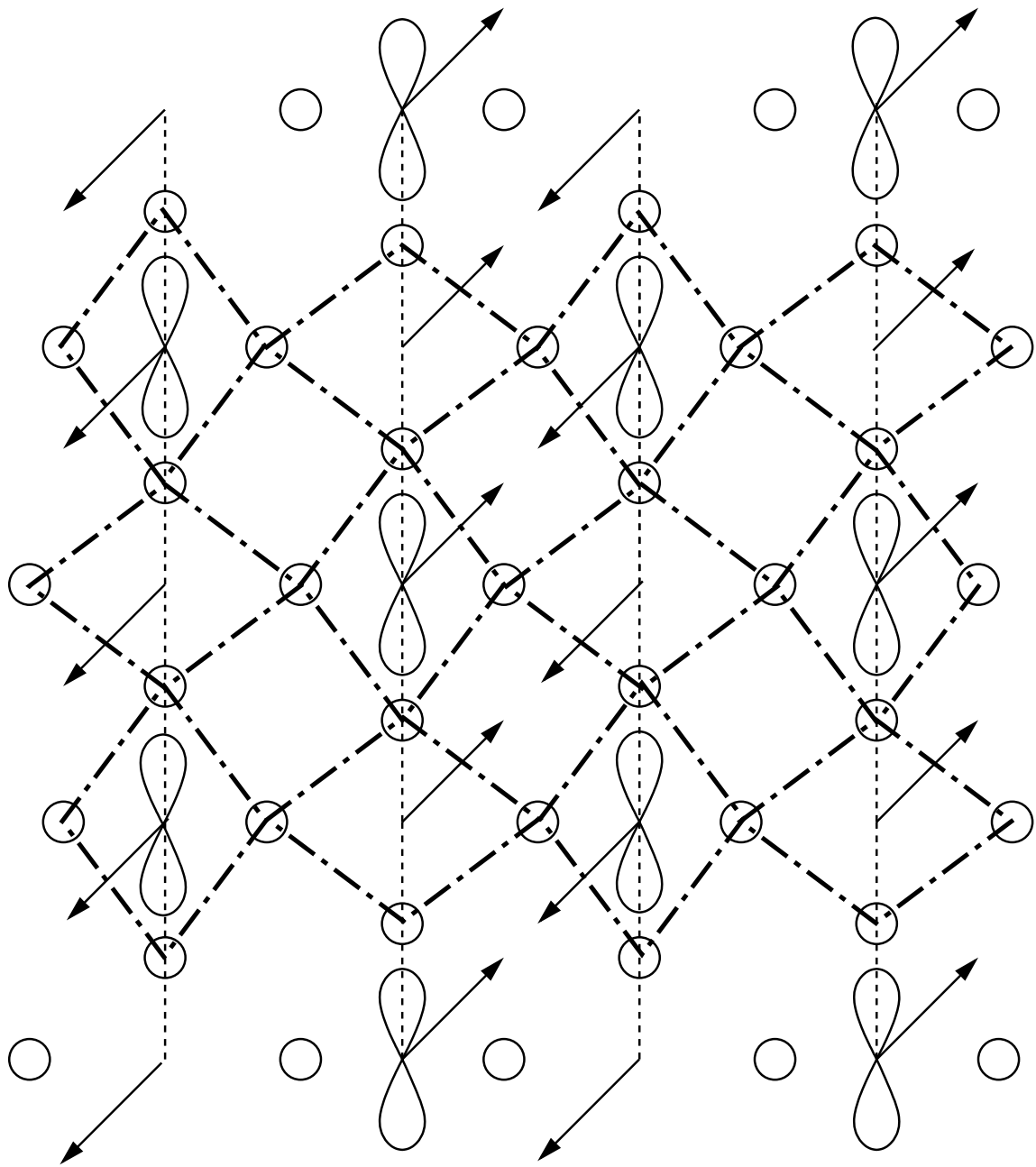
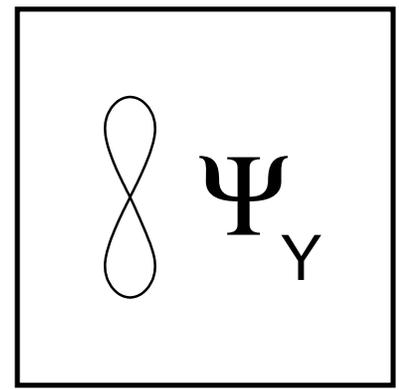
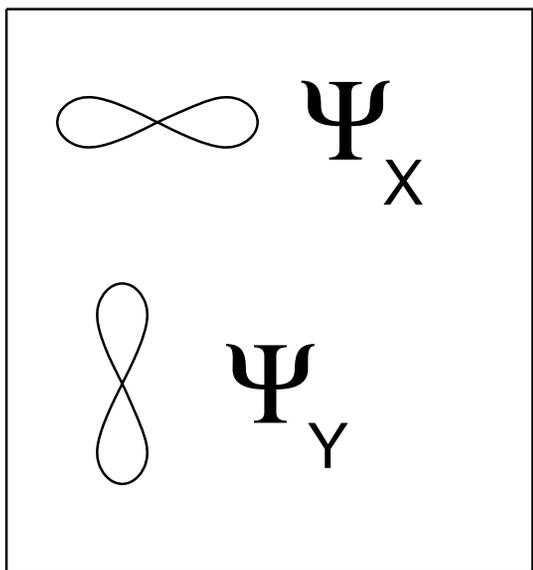
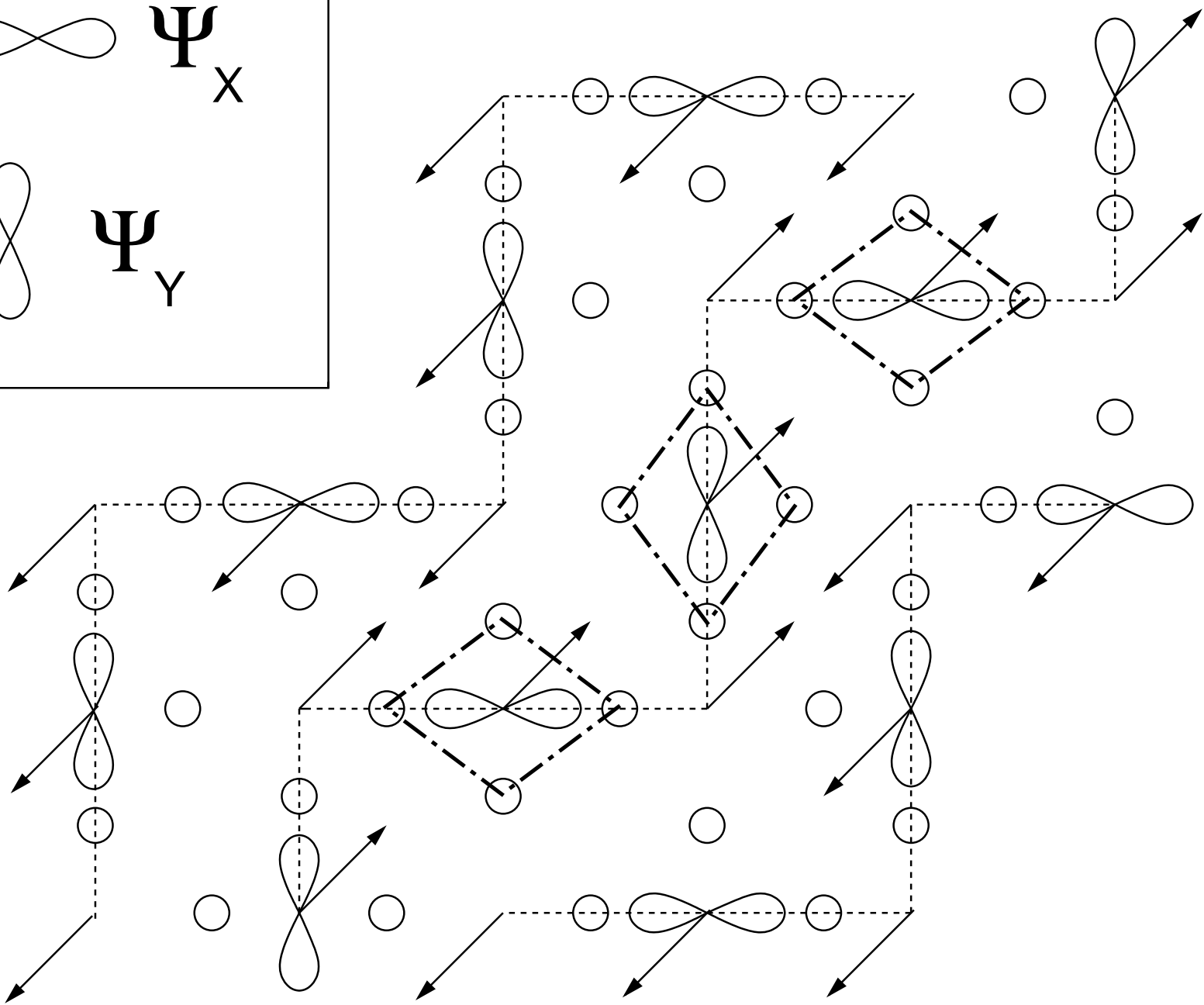
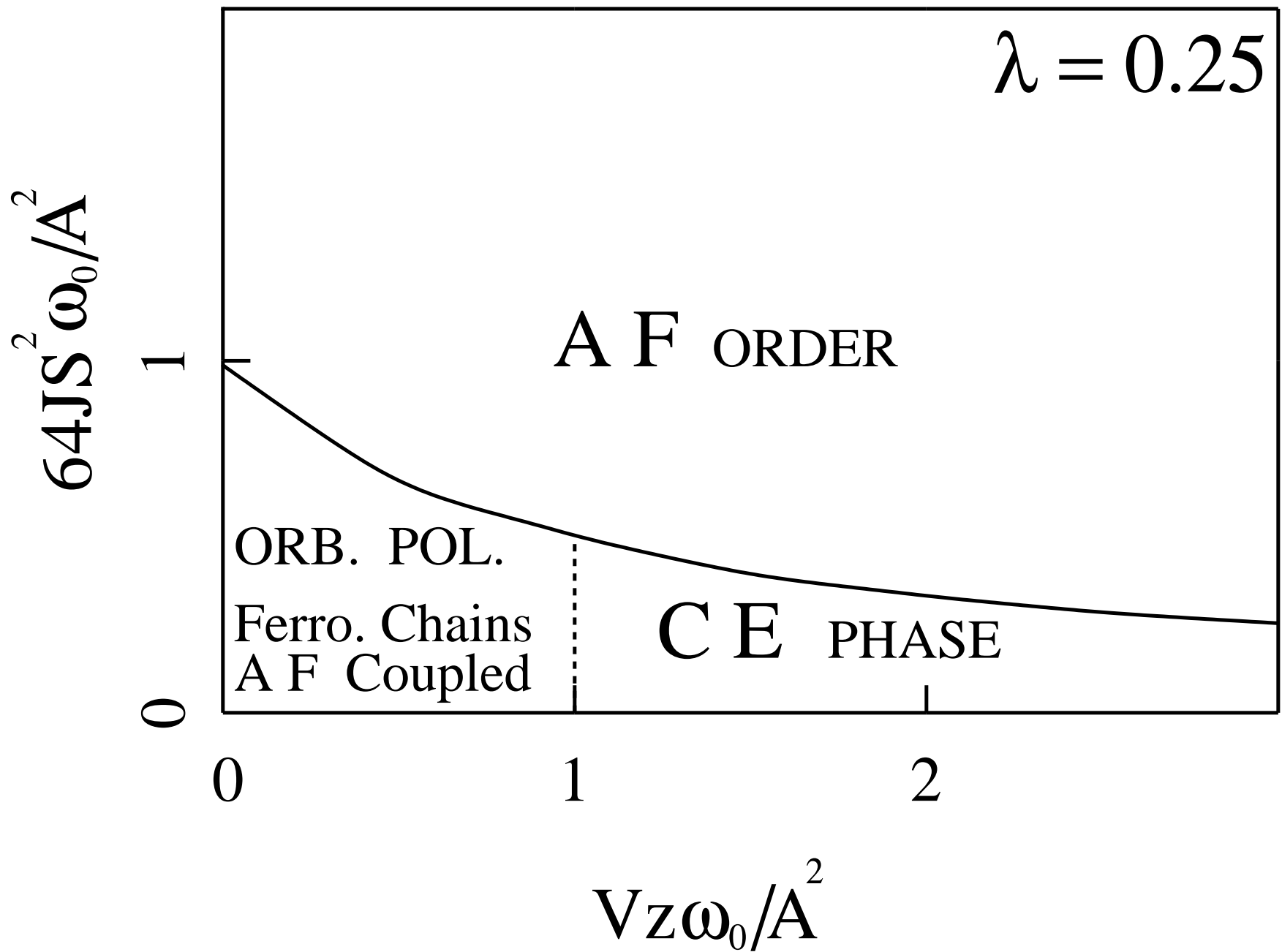
\end{document}